# Strain-mediated Symmetry Breaking Switch with Perpendicular Magnetic Anisotropy

Ruoda Zheng, Jin-Zhao Hu, Qianchang Wang, Victor Estrada, Gregory P. Carman, Abdon E. Sepulveda

*Department of Mechanical and Aerospace Engineering, University of California, Los Angeles, California, 90095, USA*

Abstract:

Magnetic switch with perpendicular magnetic anisotropy (PMA) is a promising method for controlling magnetization in several applications like magnetic tunnel junction and magnetic memory. However, incoherence happens during the switch process and lower the switch frequency of the magnetic bits. Symmetry broking can help solve this problem. Here, we present a field-free method for the symmetry broking and then increase the switch speed of the magnetization. A strain-mediated method with geometric asymmetry is presented here. In this work, we build a finite element model that consists a 50 nanometer diameter nanodisk with a varied thickness on the top of a 50 nanometer thick PZT (Pb$_y$[Zr$_x$Ti$_{1-x}$]O$_3$) thin film. The results show a 66% faster switch than symmetry PMA switching (0.85 nanosecond to 0.29 nanosecond) under same energy consumption. Finally, we explore the mechanism of the symmetry broking of varying thickness nanodot with calculating the energy profile.

In the last couple of decades, considerable research has been conducted on how to perform magnetization switching in the nanoscale due to its promise for applications in magnetic memory and magnetic logic. Switching has previously been achieved using current-induced spin transfer torque (STT), spin orbit torque (SOT) [1–11] However, a successful switch using these methods is usually accompanied by substantial resistive heat losses resulting in inefficiencies. For this reason, eliminating the need for electric current in switching methods is of great importance. An alternative to this, is using voltage-controlled methods. A promising example of such, is using strain-mediated multiferroic composites (ferromagnetic–ferroelectric heterostructure) due to their ultra-high energy efficiency.[12,21,22,13–20] Wang et al. have shown that using strain-mediated multiferroic composites with perpendicular magnetic anisotropy (PMA) can be used to achieve 180 degree magnetization switching .[23,24] However, these structures have limited switching frequencies and there are very few approaches that can be used to increase their performance. The most commonly used method is increasing the input voltage to the system, but this sacrifices the energy efficiency of these strain-mediated multiferroic structures.

One of the main reasons for the limitation of the switching speed in strain-mediated PMA structures is the incoherence at the beginning of the switching process. Some portions of the magnet switch before the others. In order to avoid this, the breaking of structural symmetries in the nanomagnetic system has attracted substantial interest. This symmetry-breaking gives the magnetization a preferable switching direction. Previous researchers have demonstrated several methods for breaking this magnetization switch symmetry. [1,25,26] One approach is using a sloped surface to help break the structural symmetry and this has been shown to work experimentally. [1] However, there is currently little simulation work done on this symmetry-breaking for 180-degree PMA switching. In addition, most simulations previously done on micromagnetics systems only consider one-way coupling from mechanical strain to magnetization.[27–29] To address this, we've created a model that reflects the influence of magnetization change on strain distribution. This change makes noticeable differences in the magnetoelastic behavior of materials.

In this paper, a finite element model is presented to simulate the strain-mediated magnetization switch in a sloped nanodisk with PMA. It is demonstrated that the switch speed is 65% faster than that of a traditional PMA nanodot while remaining equally efficient. Also, an energy profile analysis of different slope geometries was performed to show that the use of a sloped surface indeed breaks the symmetry. The comparison between a planar nanodisk and a sloped nanodisk shows how breaking the structural symmetry can help reduce the incoherence during magnetization out-of-plane (OOP) switching and improve the switching speed.

For the fully coupled model, linear piezoelectric and linear elastic behavior are assumed in the electrostatic limit. Additionally, the mechanical loss of the thin film piezoelectric transducer (PZT-5H) is assumed negligible.[30] This assumption is reasonable because, in practice, highly dense PZT fabricated by proper methods can be demonstrated to operate at ultra-high frequencies with an acceptable loss factor.[31] Alternatively, higher loss factors can be overcome with larger applied voltages or domain engineered piezoelectric/electrostrictive materials.[28] The elastic strain $\boldsymbol{\varepsilon}_{el}$ and electric displacement **D** are determined from the following coupled constitutive equations:

$$\boldsymbol{\varepsilon}_{el} = s_E \cdot \boldsymbol{\sigma} + d^t \cdot \mathbf{E} \tag{Eq. 1a}$$

$$\mathbf{D} = d \cdot \boldsymbol{\sigma} + e_\sigma \cdot \mathbf{E} \tag{Eq. 1b}$$

where $\boldsymbol{\sigma}$ is the stress, **E** is the electric field, $s_E$ is the piezoelectric compliance matrix measured under constant electric fields, $d$ and $d^t$ are the piezoelectric coupling matrix and its transpose, and $e_\sigma$ is the electric permittivity matrix measured under constant stress.

Thermal fluctuations are ignored for modeling the magnetization dynamics of the magnetoelastic dot.[32–36] Also, the magneto crystalline anisotropy of the material is neglected due to the magnetic film deposited being amorphous. The precessional magnetization dynamics of the single domain magnetoelastic dot are governed by the Landau-Lifshitz-Gilbert (LLG) equation:

$$\frac{\partial \mathbf{m}}{\partial t} = -\mu_0 \gamma (\mathbf{m} \times \mathbf{H}_{eff}) + \alpha_G \left( \mathbf{m} \times \frac{\partial \mathbf{m}}{\partial t} \right) \tag{Eq. 2}$$

where **m** is the normalized magnetization, $\mu_0$ is the vacuum permittivity, $\gamma$ is the gyromagnetic ratio (~1.76e11 Hz/T[6]), $\mathbf{H}_{eff}$ is the effective magnetic field, and $\alpha_G$ is the Gilbert damping parameter. The first term on the RHS represents precessional torque accounting for the gyromagnetic motion while the second term represents damping of the precessional motion. The effective magnetic field can be obtained from the total energy density and is given by:

$$\mathbf{H}_{eff} = -\frac{1}{\mu_0 M_s} \frac{\partial E_{tot}}{\partial \mathbf{m}} = \mathbf{H}_{ext} + \mathbf{H}_{ex} + \mathbf{H}_{anis} + \mathbf{H}_d + \mathbf{H}_{me}(\mathbf{m}, \boldsymbol{\varepsilon}) \tag{Eq. 3}$$

where $E_{tot}$ is the total energy density, $\mathbf{H}_{ext}$ is the applied external magnetic field, $\mathbf{H}_{ex}$ is the effective exchange field, $\mathbf{H}_{anis}$ is the effective anisotropy field, $\mathbf{H}_d$ is the effective demagnetization field, and $\mathbf{H}_{me}(\mathbf{m}, \boldsymbol{\varepsilon})$ is the effective magnetoelastic field. The effective magnetoelastic field is dependent on the total strain ($\boldsymbol{\varepsilon}_{el}$) which is the sum of the elastic strain ($\boldsymbol{\varepsilon}_{el}$-Eq. 1a) and magnetoelastic strain ($\boldsymbol{\varepsilon}_m$). For amorphous materials, the magnetoelastic strains are given by:

$$\varepsilon_{ij}^m = 1.5\lambda_s \left( m_i m_j - \frac{1}{3} \right), \quad i = j \tag{Eq. 4a}$$

$$\varepsilon_{ij}^m = 1.5\lambda_s m_i m_j, \quad i \neq j \tag{Eq. 4b}$$

where the saturation magnetostriction $\lambda_s$ is a material constant.[37]

The fully coupled model includes mechanically free boundary conditions for the patterned electrodes and top surface of the PZT layer. The bottom surface is fixed to simulate clamping of the PZT thin film grown on a Si/Pt substrate. To simulate a device fabricated on a larger substrate, roller boundary conditions are placed on all sidewalls. These boundary conditions are chosen to simulate a fabricated device.

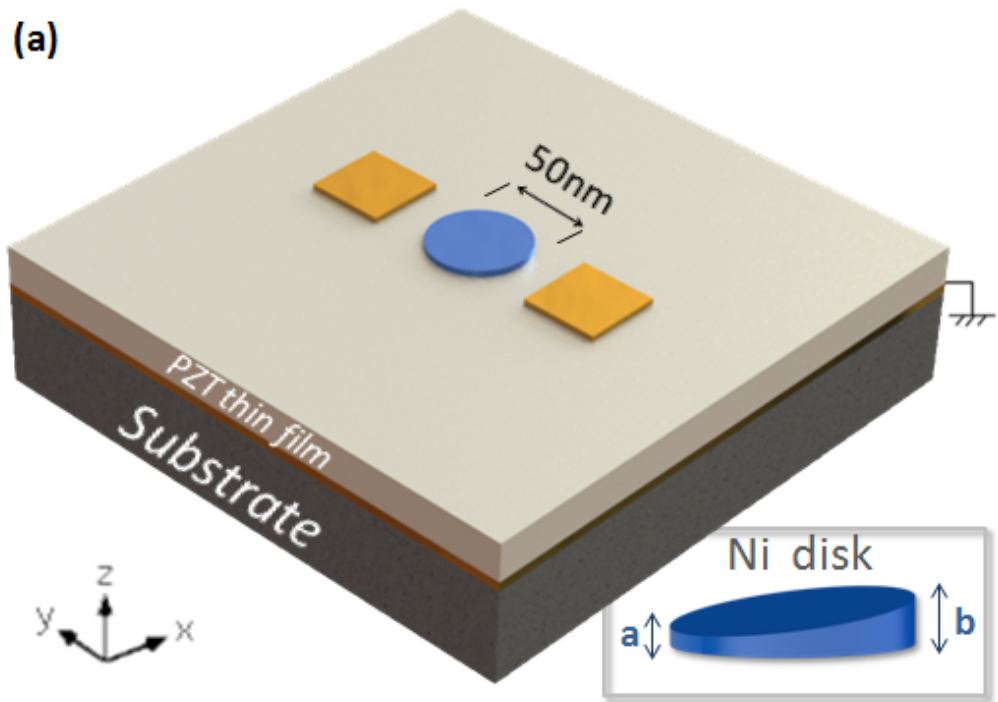

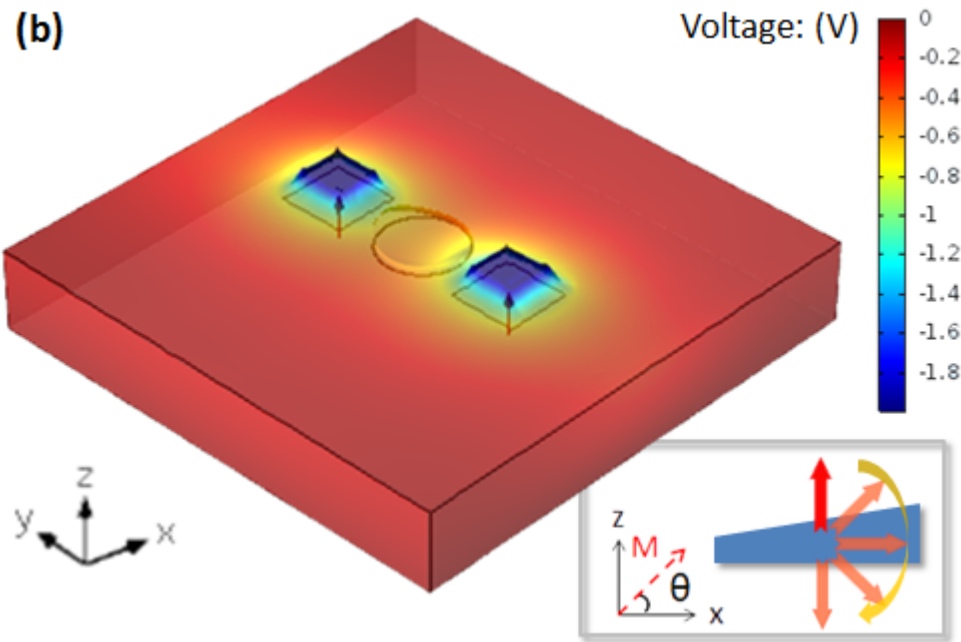

Figure 1: (a) Model setup illustration and schematics of the Ni disk. (b) Voltage value with deformation and schematics of 180° precessional magnetization switching

The Ni/PZT island structure of the MTJ system is shown schematically in Figure 1(a) including a Ni disk with non-uniform thickness. The diameter of the disk is fixed at 50nm, and the thickness varies from a (2.5 nm) to b (3.5 nm). The local enlarged image of the disk shows that the thickness varies along $\hat{x}$. The nickel disk is adhesively attached to the PZT thin film. This disk is originally magnetically pre-poled along $\hat{z}$. Beside the Ni disk, two Au electrodes (50nm×50nm) are located 20 nanometers away from the edge of the disk. The bottom of the PZT film is fixed to a thick Si substrate and electrically grounded. At the same time, roller boundary conditions were applied on the side walls of the PZT which limits their displacement to only the vertical direction.

Figure 1(b) shows the deformation of the electrodes and the Ni disk when negative voltage is applied on the electrodes. The color bar shows the voltage value (V) and the arrows underneath the electrodes indicate the direction of the electrical field: from bottom to up along the $\hat{z}$ axis. This deformation induces tension along $\hat{y}$ and compression along $\hat{x}$ within the nanondisk.[35] Due to the negative $\lambda_s$ and small Gilbert damping of Ni, the magnetization will rotate along $\hat{x}$ and overshoot. By then releasing the voltage at a proper time, 180° magnetic switching can be achieved within the nanodisk, as shown in the inset. As shown in the cross-section view of the Ni nanodisk, the big red arrow indicates the volume average magnetization. If we apply and release the voltage properly, the volume average magnetization will rotate 180 degree. The angle between the volume average magnetization and the $\hat{x}$ axis is defined as $\theta$. As described previously the thickness of the Ni disk is set to 2.5nm-3.5nm originally and can be modified as long as the energy barrier and PMA effects conditions are satisfied. Specifically, the energy barrier should be larger than $40k_BT$ (about 0.2 aJ) to avoid spontaneous magnetization switching and the thickness should stay in proper range (about 1.5nm-6nm) where the PMA effect dominates.[24] The parameters used in this model are presented in the following Table 1[23,24]:

Table 1: Material parameters

| Parameter | Description | Value |
| --- | --- | --- |
| $M_s$ | Saturation magnetization | $4.8 \times 10^5$ (A/m) |
| $\alpha$ | Gilbert damping factor | 0.038 |
| $A_{ex}$ | Exchange stiffness | $1.05 \times 10^{-11}$ (J/m) |
| $\lambda_s$ | Saturation magnetostriction coefficient | $-34 \times 10^{-6}$ |
| E | Young's modulus | $219 \times 10^9$ (Pa) |
| $\rho$ | Density | 8900(kg/m^3) |
| $\upsilon$ | Poisson's ratio | 0.31 |

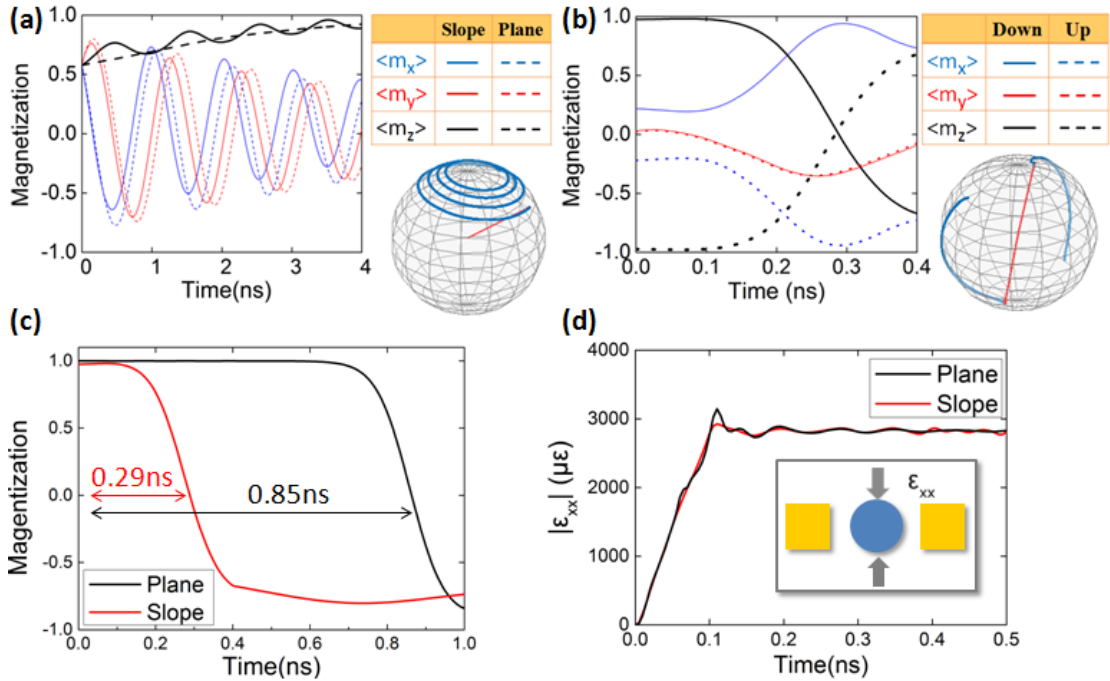

Figure 2: (a) Free precession to equilibrium state after releasing from (1, 1, 1) direction of planar and sloped nanodots. (b) Switching up and down process of the sloped nanodot with applied voltage. (c) Switching time comparison of planar and sloped disks when

Figure 2(a) shows the dynamic response of the normalized magnetization of sloped and planar Ni disks after releasing it from the $(1, 1, 1)$ direction. The solid line represents the dynamic response of the sloped disk while the dashed line represents the planar one. As shown in the figure, the magnetization components $m_x$, $m_y$, $m_z$ process towards the $\hat{z}$ axis with $m_x$, $m_y$ oscillating. Finally, the magnetization reaches an equilibrium position in the $\hat{z}$ direction. Both the sloped and planar disk equilibrium processes are similar with the one significant difference being that the $m_z$ of the sloped disk does not increase as monotonically as the planar one. In the global plot, the trace of the volume average magnetization for the sloped disk is shown. Figure 2(b) illustrates the dynamic response of the magnetization for a sloped Ni disk after an electric field is applied. The solid line represents the dynamic response from $\hat{z}$ to $-\hat{z}$ and the dotted line represents the response from $-\hat{z}$ to $\hat{z}$. Again, the global plot shows the trace of the volume average magnetization for the sloped disk. The electric field $E_c$ is applied after the magnetization reaches equilibrium, which is the 0 ns state specifically. The duration of the 40MV/m electrical field application is 0.4ns.

The relaxation time RC of PZT film (0.2ps) and the mechanical strain propagation time from the electrodes to the disk (4.0ps) can be assumed negligible. Evidently, for the planar nanodisk, the $m_z$ component of two switching processes is symmetric due to the geometry. However, for the sloped Ni nanodisk, $m_x$ is antisymmetric, indicating that the magnetization always switches clockwise. This is due to the fact that the sloped disk results in broken structural symmetry about $\hat{x}$ axis. Figure 2(c) shows that, under similar strain, the magnetization switching time varies significantly. A large reduction of the switching time is observed in the sloped disk when compared to the planar disk. The planar disk takes 0.85ns (1176MHz) for the volume average magnetization to become in-plane ($m_z = 0$) while the sloped one takes only 0.29ns (3448MHz and 65% faster). Traditionally, for the planar disk, $m_z$ changes little before 0.6ns due to the incoherence of the magnetization switching. In other words, in the planar disk the magnetization hesitates to choose a direction of switching. To increase the switching speed higher voltages would have to be applied.[33] However, for the sloped disk case, the PMA effect in the thicker part is weaker. So that all magnetization spins switch together toward the thicker part of the nanodisk and thus respond much faster. Figure 2(d) shows that, although the nanodot dimension is smaller than the approximate single-domain limit of Nickel ($L_{SD} = 10\sqrt{\frac{2A_{ex}}{\mu_0 M_s^2}} = 85$nm), strain inside the element still vary with time at the beginning of the switching, indicating non-uniform strain distribution and incoherent switching which reduce the energy efficiency. However, by breaking the shape symmetry, the sloped nanodot can provide a solution for the dilemma between fast reaction speed and coherence of switching.

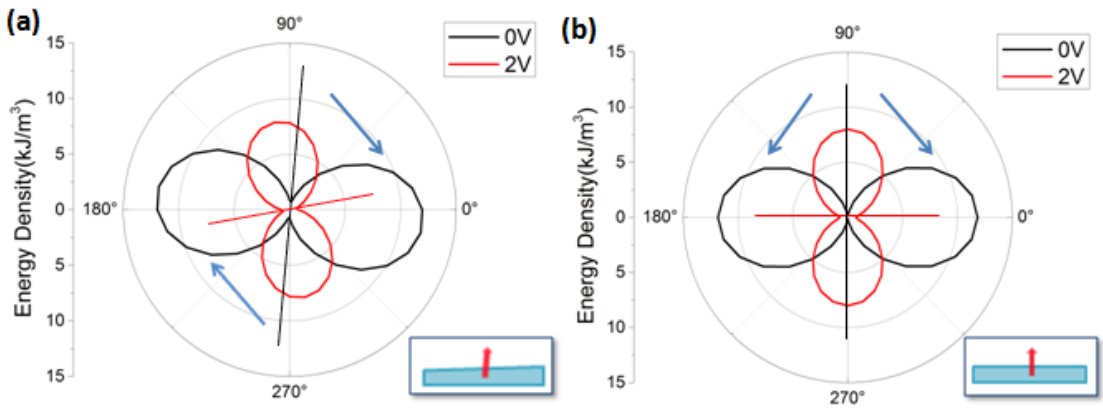

Figure 3: *(a) Energy profiles in x-z plane of the sloped nanodot before and after the voltage is applied. Black and red lines indicate EA correspondingly. (b) Energy profiles in x-z plane of the*

*planar nanodot.*

Figure 3 shows the energy profile of sloped and planar elements before and after the voltage is applied. It can provide an explanation for the deterministic rotation from the perspective of energy. From Figure 3 (a), the free easy axis (EA) of the sloped disk is along $\theta = 84°$ and the EA with voltage applied is along $\theta = 10°$. The hard axis (HA) is always perpendicular to the EA in each case. This property induced by creating shape anisotropy is a little different from the planar disk whose EA lies along exactly at $\theta = 90°$ and $\theta = 0°$ before and after the magnetization rotation respectively, as can be seen in Figure 3(b). The energy asymmetry can determine the magnetization rotating direction. For example, starting from $\hat{z}$, when the strain is introduced, the volume average magnetization tends to reorient along positive $\hat{x}$ because reaching negative $\hat{x}$ will require more energy. Similarly, it rotates from $-\hat{z}$ to $\hat{z}$ clockwise. But for the planar disk, rotating through positive $\hat{x}$ or negative $\hat{x}$ needs equal energy. Therefore, incoherence emerges, part of the spins tends to rotate through one direction while others tend to rotate though other direction, and the reaction time is prolonged.

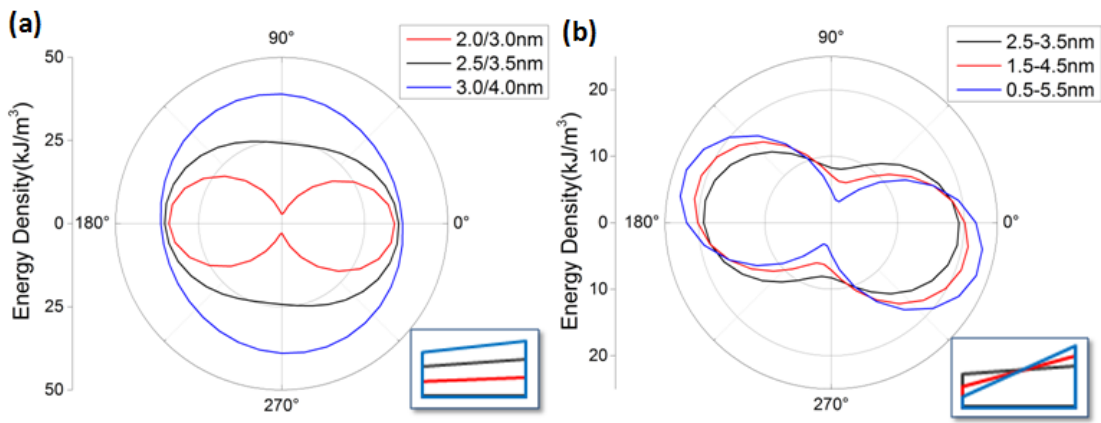

Figure 4: Free energy profiles of nanodots with different thickness and inclination.

So far, the article has focused on the dynamic magnetic response of a sloped nanodot. Now we will look at the effect that changing certain geometry parameters of the nanodot (thickness and inclination) will have on the overall performance. Figure 4 shows the energy profiles of sloped nanodots with different thicknesses and inclinations. The nanodot schematics are shown in the insets. Setting average thickness to 2.5nm/3nm/3.5nm with a specified height difference value of 1nm between the highest and lowest points, the energy profiles' shape looks different as seen in Figure 4 (a). It is important to note that the energy profile shape itself can be changed by adjusting the zero potential energy position. However, putting them in the same plot, we realize that the energy barrier is decreasing when the sloped nanodot becomes thicker. This tendency matches the former conclusion that the PMA effect is inversely proportional to thickness. The magnetostatic energy will drag the magnetization EA from out-of-plane to in-plane at thickness greater than 6nm in Ni nanodots. The EA direction also changes but is negligible compared to the dramatic change of energy barrier. However, for sloped nanodots with different inclinations, the change of EA direction dominates and has a distinct tendency. In Figure 4(b), the difference value of a and b is set to 1nm/3nm/5nm with the average thickness remaining at 3nm. The EA directions corresponding these inclinations are $\theta = 84°$, $\theta = 76°$, $\theta = 74°$. It is evident that the magnetization tends to become in-plane as the slope of the nanodots become larger. The tendency shown in the data above suggests that the degree of deterministic rotation can be strengthened by breaking the symmetry more drastically. Nonetheless, sharply sloped nanodots are not required to achieve deterministic rotation. Little inclination combined with mild voltage can avoid incoherence, saving both switching time and wasted energy.\

The simulation results demonstrated that under voltage induced strain the magnetization can archive 180-degree OOP switch in Nickel nanodisk. When the symmetry is broken, the switch speed can increase 66% according to the simulation results. Under 40MV/m electrical field, the switch time for plane nanodisk is 0.85ns while for the sloped nanodisk is 0.29ns. This work avoids the trade-off between energy efficiency and coherence of magnetization as traditional way has. Results validate the feasibility that use geometric way to break the symmetry and then avoid the incoherence. However, additional work is required for developing a easy way for fabricatng the system shown in this paper.

This work was supported by NSF Nanosystems Engineering Research Center for Translational Applications of Nanoscale Multiferroic Systems (TANMS) Cooperative Agreement Award No. EEC-1160504. R. Z. and J.-Z. H. contribute equally to this work. J.-Z. H. conceived the design and R. Z. led the studies. R. Z. and J.-Z. H. performed the finite element simulations. R. Z. and